\pdfoutput=1

\documentclass[10pt, preprint2]{aastex}

\begin{document}

\title{Sensitive Observations of Radio Recombination Lines in Orion and W51: The Data and Detection of Systematic Recombination Line Blueshifts Proportional to Impact Broadening}

\shorttitle{Sensitive Observations of Radio Recombination Lines in Orion and W51}

\author{M.B. Bell\altaffilmark{1}, L.W. Avery\altaffilmark{2}, J.M. MacLeod\altaffilmark{2}, J.P. Vall$\acute{e}$e\altaffilmark{2}}

\altaffiltext{1}{Herzberg Institute of Astrophysics,
National Research Council of Canada, 100 Sussex Drive, Ottawa,
ON, Canada K1A 0R6; morley.bell@nrc-cnrc.gc.ca}

\altaffiltext{2}{Herzberg Institute of Astrophysics,
National Research Council of Canada, 5071 West Saanich Road, Victoria, BC V9E 2E7} 

 \begin{abstract}

Sensitive spectral observations made in two frequency bands near 6.0 and 17.6 GHz are described for Orion and W51. Using frequency switching we were able to achieve a dynamic range in excess of 10,000 without fitting sinusoidal or polynomial baselines. This enabled us to detect lines as weak as T$_{A} \sim$ 1mK in these strong continuum sources. Hydrogen recombination lines with $\Delta$$n$ as high as 25 have been detected in Orion. In the Orion data, where the lines are stronger, we have also detected a systematic shift in the line center frequencies proportional to linewidth that cannot be explained by normal optical depth effects.

\end{abstract}
 
\keywords{HII regions--- ISM: abundances --- radio lines: ISM}

\section{Introduction}

In a recent paper \citet{gor08} has discussed how radio recombination lines can be useful tools for astronomers and physicists. Because of their high spectral density, observations in a narrow frequency band typically encompass lines corresponding to a wide range of $n$ values, where $n$ is the principal quantum number. Provided the sensitivity is high enough, such observations allow the study of impact broadening and source analysis based upon many transitions, free of systematic errors due to differences in beam sizes, calibration errors and telescope efficiencies. Another advantage of observing lines of different $n$ in the same window is that the optical depth of the radio continuum is the same for all lines \citep{chu71,loc75,pei92}. With the same absorption effects operating in each line it is possible to make a more reliable investigation of line broadening mechanisms such as turbulence and electron impact broadening \citep{gri67}.
     
We report here the results obtained by making small-offset, frequency-switched observations (hereafter referred to as SOFS observations) at a fixed frequency, with a relatively wide, instantaneous bandwidths (140 MHz at 17.6 GHz and 70 MHz at 6.0 GHz), in W51 and Orion A, two sources with strong continuua. The source coordinates used were R.A.(1950) = 05$^{h}$32$^{m}$50$^{s}$; Dec(1950) = -05$\arcdeg$25$\arcmin$l2$\arcsec$ for Orion A and R.A.(1950) = 19$^{h}$21$^{m}$26$^{s}$; Dec(1950) = 14$\arcdeg$24$\arcmin$40$\arcsec$ for W51(Main). Our observations were carried out using the NRAO\footnotemark 140-foot (43 m) telescope. The data at 6.0 GHz were obtained in 1990 August (W51) and 1992 April (Orion); the data at 17.6 GHz in 1991 January.

\footnotetext[3]{The National Radio Astronomy Observatory (NRAO) is operated by Associated Universities Inc., under agreement with the National Science Foundation.}

We present these data for publication at this time for two main reasons. First, in the ten years since we first described our observing and data reduction techniques \citep{bel97}, and reported a mysterious reduction in impact broadening above $n$ = 200 \citep{bel00}, we have had many requests from astronomers asking to see the data. Second, we fully anticipated that when the Green Bank Telescope (GBT) was commissioned ten years ago, it would soon be used either to confirm or to rule out our earlier findings. This has not happened. To our knowledge, even the GBT, with its offset feed designed to reduce the feed-to-primary dish reflection problem, and its extensive receiver-room shielding to control local interference problems, has not obtained position-switched results over wide windows close to the 10,000/1 dynamic range we achieved with the 140-foot telescope using frequency switching. We hope by presenting our data it might encourage someone to use the GBT to try to explain the mysterious line narrowing problem above quantum levels of $n$ = 200 that we previously reported.  


Here, the spectra obtained for each source are presented at both observing frequencies. The peak antenna temperatures and linewidths (FWHM) are tabulated. For sources with a strong continuum, frequency switching offers the opportunity to observe baseline-free spectral windows that are many times wider than the continuous window widths obtainable using position switching. Although the same correlator window widths can be used with both techniques, because of the need to fit high-order polynomials to position-switched data, only small portions of the total observing window can be processed at a time.

Observations have been reported previously of recombination lines arising from transitions between levels separated by $\Delta$$n$ up to at least 14 \citep{bal94,roo84,ban87,bel95,dav71,smi84}. Here we report recombination lines with $\Delta$$n$ values possibly as high as $\Delta$$n$ = 25. Throughout this paper the principal quantum number $n$ represents the lower level of a given transition whose quantum designation is given as ($n,\Delta$$n$). 

In \S 2 we describe the receivers and observing technique. In \S 3.1 we describe the spectra obtained for both sources at 17.6 GHz, including molecular lines. In \S 3.2 the spectra obtained at 6 GHz are described. In \S 3.3 some of the main molecular lines in our observing window are discussed. In \S 4 we describe the RRLs listed in Tables 2, 3, 4, 6, and other, non-recombination lines listed in Table 7, as well as how the corrected values have been obtained. In \S 5 the center frequencies of the RRLs in Orion at 6 GHz are examined and shown to exhibit a systematic blueshift as $n$ and $\Delta n$ increase. A brief description of the correction curves used to recover the true line parameters is included in Appendix A.

\section{Observations}

As reported by \citet{bel91}, it turned out to be necessary to make off-source integrations of equal duration, at least for the first several hours to ensure that the spectra were not contaminated by telluric lines or interference. Using frequency-switching, these lines, that otherwise disappear with off-source subtraction, are readily detected. However, this possible spectral contamination is not too serious since telluric lines can usually be identified, and even time-variable weak interference can be discovered by dividing the data into two halves and subtracting them. All results were obtained using relatively short, 4-minute ON-OFF integration periods. To reduce the fundamental telescope ripple we also used $\lambda$/8 defocussing.

The autocorrelator was configured to give two series banks of 80 MHz and 256 channels at 17.6 GHz which gave channel separations of 0.3125 MHz or 5.32 km s$^{-1}$. The bank center frequencies were separated by 70 MHz, allowing a frequency overlap of 10 MHz to accommodate autocorrelator edge effects. The observing window thus covered 140 MHz, and with cold-sky system temperatures between 50-60K at 17.6 GHz the desired sensitivity was obtained in six transits.   Telescope pointing was updated every 90 minutes using standard pointing sources. The relevant telescope beamwidths and efficiencies are given in Table 1.

At 6.0 GHz the autocorrelator was configured to give two series banks of 40 MHz width and 512 channels which gave channel separations of 0.0781 MHz or 3.9 km s$^{-1}$. The frequency offsets used at 17.6 and 6.0 GHz were $\pm0.9375$ MHz and ±0.3125 respectively, (which correspond.to frequency-switched intervals of 32 and 31.2 km s$^{-1}$). Although a slightly larger offset would have been desirable, the baseline degradation with increasing offset ruled this out. The flatness of the baseline and hence the line-detecting ability of the observing technique was felt to be more important than the loss of line parameter information which may have resulted, especially since the latter can be recovered. Also, removing the frequency-switched reference lines so that Gaussians can be fitted requires that the baselines be flat.

The LINECLEAN program used in the reduction process employed a scaled version of the signal line to remove the reference lines. It is described in detail in \citet{bel97}. In the reduction process the known frequencies of all recombination lines, up to $\Delta$$n$ = 20 at 17.6 GHz and up to $\Delta$$n$ = 25 at 6 GHz, were examined and were cleaned if they showed evidence of a line. Very weak features that were confused with other features were not cleaned. A complete discussion of the procedures used to obtain flat baselines with the SOFS observing technique, without having to fit polynomial or sinusoidal baselines, is given in \citet{bel97}.

\section{Results}

\subsection{The 17.6 GHz data}

The spectrum of Orion A near 17.6 GHz is shown in Figure 1 on a scale that reveals the weakest lines.  No sinewaves or polynomials have been removed from this spectrum even though the continuum brightness temperature of the source is in excess of 40K. A list of the detected recombination lines is presented in Table 2. It is immediately apparent in Figure 1 that the baseline structure, even at the 1 mK level, is limited mainly by thermal noise. The only obvious non-recombination line, detected in this window, that is of reasonable width, is at 17595.3 MHz. This corresponds to the overlapping 38$_{16,23}$ - 37$_{17,20}$ and 38$_{16,22}$ - 37$_{17,21}$ transitions of CH$_{3}$COOH which is a common molecule in Orion. The 2p$\sigma_{\mu}$(V=0,N=2,G=l/2) - ls$\sigma_{g}$(V=19,N=l,G=l/2) and 2p$\sigma_{\mu}$(V=0,N=2,G=3/2) -  ls$\sigma_{g}$(V=19,N=l,G=3/2) transitions of H$_{2}^{+}$ at 17604.3$\pm±0.5$ and 17610.3$\pm±0.5$ GHz respectively \citep{car93}, also fall inside this window and were one of the main reasons for its choice. These results provide a very sensitive upper limit to the amount of H$_{2}^{+}$ in Orion A. Although features at these frequencies might be confused with the H(163,13) line at 17606.8 MHz, any H$_{2}^{+}$ line present is unlikely to be stronger than T$_{A} \sim 3$mK. 

    
The 17.6 GHz spectrum of W51 is shown in Figure 2 and a list of the detected lines is given in Table 3. Although the spectrum is similar in appearance to Orion A, the recombination lines are roughly 50\% weaker. As in the Orion A spectrum, there is no evidence for H$_{2}^{+}$ to a level of $\sim 1$ mK.

\subsection{The 6 GHz data}

The spectrum of Orion A at 6 GHz is presented in Figure 3 with the vertical scale adjusted so that a line of strength T$_{A}$ = 1 mK is visible. The sensitivity is significantly better in the high-frequency half of the spectrum due to a change in receiver noise across the 70 MHz bandwidth. The locations of all recombination lines with $\Delta$$n < 25$ are indicated whether or not a line has been detected. There is a strong, variable feature near 6005 MHz that has prevented the detection of weak lines within at least 2 MHz to either side.

Although it was assumed to be interference, it is worth noting that the $^{2}\Pi_{3/2}$ (2.5$_{1.5,1.0}$ - 2.5$_{0.5,0.0}$) transition of $^{17}$OH is located at 6005.6 MHz and the possibility that this line may be masering cannot be ruled out. A list of the lines detected is given in Table 4. The notes to Table 4 are given in Table 5. Although 41 lines are listed, only the rest frequency is given if the line is confused with a stronger one. About ten percent of the lines appear to be molecular in origin and some of these are discussed in more detail in \S 3.3 below and in Fig 6. The H(191,10) line near 8 GHz, was detected by \citet{roo84}, with T$_{A}$ = 6 mK. For comparison purposes the $\Delta$$n$ = 9 line at 6 GHz in Orion reported here in Table 4, has corrected peak antenna temperature of T$_{A}$= 12.7 mK (see \S 4 for a brief description of how the corrected peak antenna temperatures are obtained). It is not immediately obvious why the H(213,14) line near 8664 MHz was not visible in the spectrum of Orion obtained by \citet{ban87}, or why they did not detect the H(203,12) or H(208,13) lines, especially in M17S, but these failures might have resulted from the fact that their data required the removal of a 12th order polynomial from the baseline.

The spectrum of W51 obtained at 6 GHz is presented in Figs 4 and 5, with the vertical scale adjusted to show the 1 mK level. Again there is no baseline structure apparent above the level of the thermal noise, except within the cleaned region of the stronger lines and no baselines have been removed. Recombination lines up to at least $\Delta$$n$ = 14 appear to be visible when they are located in unconfused regions. A list of the detected lines is given in Table 6. The corrected peak antenna temperature for the H(194,7) line is 11.5 mK in Table 6, which shows good agreement with the T$_{A}$=14.1 and 16.0 mK reported by Bania et al, (1987) for the H(171,7) and H(170,7) lines in this source, using the same telescope.

In Table 7 we have listed similar information for the Helium recombination lines detected in W51 in the 6 GHz window. Table 8 lists the frequencies and antenna temperatures of all molecular lines detected in W51 at 6 GHz.

\subsection{Molecular Lines}
 
Three components of the $^{2}\Pi_{3/2}$ transition of $^{16}$OH, (2.5$_{1.5,-1,1}$ - 2.5$_{1.5,1,3}$), (2.5$_{1.5,-1,2}$ - 2.5$_{1.5,1,2}$), and (2.5$_{1.5,-1,3}$ - 2.5$_{1.5,1,3}$), near 6016, 6030, and 6035 MHz respectively, are present in the spectrum in Fig. 3 and are found to be redshifted by $\sim 22$ km s$^{-1}$, which is the velocity of the strongest outflow component near IRc2 (see e.g. \citet{han83}. The weakest falls near the H(217,10) line.

There are also 3 Q-Branch transitions of NO in this spectrum, (18.5$_{0.5,-1.0,19.5}$ - 18.5$_{0.5,1.0,19.5}$), (19.5$_{0.5,1.0,18.5}$ - 19.5$_{0.5,-1.0,18.5}$), and (19.5$_{0.5,1.0,18.5}$ - 19.5$_{0.5,-1.0,19.5}$). Although the last of these is coincident with the OH component near 6035 MHz, the remaining two fall in relatively unconfused regions near 5981 and 5993 MHz (see Table 4) and both show weak emission, also redshifted by an amount similar to that found for OH.
 
The three lines of OH are also present in Figure 5 and are considerably narrower than the recombination lines, and unresolved. The weakest OH line at 6016.7 MHz is seen in absorption and falls on top of the H(217,10) line. The two stronger OH lines, near 6030 and 6035 MHz, show both emission and absorption with the absorption component occurring in the red wing of the main emission feature. This can be seen more clearly in Figure 6 which shows an expanded view of the spectrum in the vicinity of the two strongest OH lines. The H(184,6) line in this spectrum may also show evidence for structure in its red wing, although this needs to be confirmed.

\section{Corrected Line Parameters}
   
In Table 2 we have listed in columns 1-3 all the recombination lines in Orion A with $\Delta$$n < 20$ that fall in the 17.6 GHz window, along with the H$\alpha$ and H$\beta$ lines near 17.99 and 18.94 GHz respectively. The peak strengths and linewidths (FWHM) were determined from Gaussian fits using DRAWSPEC and have been included along with 'corrected' values. The corrected values are obtained using the correction curves in Figures 12 and 13 (see also Appendix A). These adjustments are necessary to remove the effects introduced by frequency switching with small offsets.  Similar numbers for W51 are given in Table 3. In Table 4 we list in column 1 the frequencies of all 6 GHz lines detected in Orion along with the frequencies of all the recombination lines that fall inside the observing window (for $\Delta$$n < 25$).  If the line was detected, its observed frequency is listed in column 2. Line identifications are listed in column 3. The observed antenna temperature is listed in column 4 and the antenna temperature, corrected for frequency-switched effects, is listed in column 5. The linewidth (FWHM) is included in columns 6 and 7 (frequency and velocity respectively). The corrected linewidths are given in column 8. Notes in column 9 are explained in Table 5. Table 6 lists the Hydrogen recombination lines detected in W51 and Table 7 lists the Helium recombination lines in W51. Table 8 lists the other, non-recombination lines detected in W51.

\section{Line Center Frequencies}

Figures 7, 8, 9, and 10 show lines from all relatively unconfused regions of the 6 GHz Orion spectrum that contain a recombination line ($\Delta$$n < 26$). Each spectrum has the same incremental frequency scale and the rest frequency of each line is indicated by the vertical solid line. The vertical dashed line indicates the line center frequency obtained from Gaussian fitting. For lines with $n < 200$($\Delta$$n < 10$) there is a systematic shift of the line to higher frequency (blueshift) as $n$, (or $\Delta$$n$), increases. This shift is also present in previously published data and, in Fig. 11, the differences between the observed and rest frequencies reported here are compared to line frequencies reported by others \citep{smi84,roo84} for $\Delta$$n < 10$.
Since all measurements were made near the same frequency, the shifts cannot be due to normal optical depth effects related to changes in observing wavelength.

\section{Linewidths vs $n$}
     
\citet{gri67,gri74} has shown that significant line broadening due to electron collisions can be expected in high-level Rydberg-Rydberg transitions that originate in regions of relatively high electron density.  The broadening is proportional to $n^{4}$.  Examination of Tables 4 and 5, however, shows that our measured linewidths begin to decrease again for $n > 200$. This discrepancy is discussed in more detail in \citet{bel00}, who show that this effect also appears to be present in earlier data.



\section{Conclusions}
     
We have obtained sensitive spectral line observations of W51 and Orion in narrow (70 and 140 MHz wide) frequency bands at 6.0 and 17.6 GHz. A total of 61 lines have been detected, of which about 10\% are of molecular origin, 10\% are unidentified and are assumed to have molecular carriers, and the remainder are recombination lines with $\Delta$$n$-values at least as high as 21. We have used the frequency switching observing technique to make sensitive spectral observations of weak lines in sources with a strong continuum and wide linewidths, and conclude that this can be a powerful technique, especially for spectral line surveys of strong continuum sources and observations of extended regions. Finally, we have detected a shift in the line centers for hydrogen recombination lines that is proportional to impact broadening. The shift cannot be due to optical depth effects in an expanding nebula.

\section{Acknowledgments}

We thank the staff of the 140-foot telescope for their assistance in ensuring that things ran smoothly during the observing periods. We especially wish to thank George Seielstad, Green Bank site director when these observations were carried out, for seeing that adequate time was made available to complete this program. We also thank Pierre Brault for assistance with the data reduction and figure preparation. Finally, we thank an anonymous referee for several comments that helped to improve the organization of the paper.

\clearpage

\begin{deluxetable}{lll}
\tabletypesize{\scriptsize}
\tablecaption{Telescope, Receiver and Correlator Parameters. \label{tbl-1}}
\tablewidth{0pt}
\tablehead{
\colhead{Parameter} & \colhead{17.6 GHz} & \colhead{6.0 GHz}
}

\startdata

beamwidth (arcmin) & 1.7 & 5.17 \\
$\eta_{A}$ & 0.32$\pm0.03$ & 0.52 \\
$\eta_{B}$ & 0.42$\pm0.04$ & 0.65 \\
T$_{sys}$ & 50-60K & 40-50K \\
Channel width (MHz) & 0.3125 & 0.07813 \\
Freq. sw. offset(MHz) & 0.9375 & 0.3125 \\
W51 V$_{LSR}$ assumed (km s$^{-1}$) & 60 & 60 \\
Orion V$_{LSR}$ assumed (km s$^{-1}$) & 0 & 0 \\ 

\enddata 

\end{deluxetable}


\begin{deluxetable}{llllllll}
\tabletypesize{\scriptsize}
\tablecaption{Hydrogen Recombination Lines in Orion A\tablenotemark{1} at 17.6 GHz($\Delta$$n < 20$)\tablenotemark{2}. \label{tbl-2}}
\tablewidth{0pt}

\tablehead{
\colhead{Frequency(MHz)} & \colhead{$\Delta$$n$}  &  \colhead{$n$}  &  \colhead{T$_{A}$(mK)($\pm{1}$)(Raw)} & \colhead{(Corr.)} & \colhead{FWHM(MHz)} & \colhead{kms$^{-1}$} &  \colhead{Corr.}
}

\startdata

17992.56 & 1 & 71  &  1994.2 & 1994.2  & 1.138  &  20.9  & 24.5 \\
18045.89 & 2 & 89  &  503.72 & 503.7  &  1.246  &  21.2  & 25.0 \\
17569.58 & 6 & 128  & 59.64  & 60.8  &   1.339  &  22.8  & 28.3 \\
17645.81 & 8 & 140  & 25.02  & 26.0  &  1.445  &  24.6  & 32.0 \\
17696.06 & 10 & 150 & 14.56  & 15.2  &  1.443  &  24.6  & 31.9 \\
17613.65 & 12 & 159 & 10.29  & 12.8  &  1.754  &  29.9  &  43.9 \\
17606.87 & 13 & 163 & 4.02   & 4.9   &  1.729  &  29.4  &  43.0 \\
17702.32 & 15 & 170 & 1.0    & 1.3  &  1.78   &   30.3  &  45.3 window edge \\
17588.21 & 17 & 177 & 4.05   & 6.0  &  1.872  &  31.9  &  49.4 (confused?) \\
17612.76 & 18 & 180 & confused\tablenotemark{3} & -- & -- & -- & -- \\
17601.43 & 19 & 183  & $<1.0$  &  &  & & \\

\enddata 

\tablenotetext{1}{assumes $\Delta$V(Doppler) = 24.4 km s$^{-1}$}
\tablenotetext{2}{offset width = 32 km s$^{-1}$}
\tablenotetext{3}{confused with $\Delta$$n$ = 12 line}
 
\end{deluxetable}

\begin{deluxetable}{llllllll}
\tabletypesize{\scriptsize}
\tablecaption{Hydrogen Recombination Lines in W51\tablenotemark{1} at 17.6 GHz($\Delta$$n < 20$)\tablenotemark{2}. \label{tbl-3}}
\tablewidth{0pt}

\tablehead{
\colhead{Frequency(MHz)} & \colhead{$\Delta$$n$}  &  \colhead{$n$}  &  \colhead{T$_{A}$(mK)($\pm{1}$)(Raw)} & \colhead{(Corr.)} & \colhead{FWHM(MHz)} & \colhead{kms$^{-1}$} &  \colhead{Corr.}
}

\startdata

17992.56 & 1 & 71  &  887.01 & 905.1  & 1.322  &  22.04  & 28.9 \\
18045.89 & 2 & 89  &  219.21 & 230.7  &  1.382  &  23.53  & 32.0 \\
17569.58 & 6 & 128  & 26.15  & 29.4  &   1.536  &  26.15  & 38.1 \\
17645.81 & 8 & 140  & 16.19  & 16.4  &  1.243  &  21.18  & 26.8 \\
17696.06 & 10 & 150 & 6.80  & 9.4  &  1.754  &  29.87  & 47.2 \\
17613.65 & 12 & 159 & 4.62  & 6.3  &  1.736  &  29.56  &  46.6 \\
17606.87 & 13 & 163 & 3.47   & 4.6   &  1.708  &  29.09  &  45.2 \\
17702.32 & 15 & 170 & 2.54   & 3.7  &  1.795   &   30.57  &  48.8 \\
17588.21 & 17 & 177 & 2.22   & 2.2  &  1.239  &  21.10  &  26.9  \\
17612.76 & 18 & 180 & confused\tablenotemark{3} & -- & -- & -- & -- \\
17601.43 & 19 & 183  & $<1.0$  &  &  & & \\

\enddata 

\tablenotetext{1}{assumes $\Delta$V(Doppler) = 28.9 km s$^{-1}$}
\tablenotetext{2}{offset width = 32 km s$^{-1}$}
\tablenotetext{3}{confused with $\Delta$$n$ = 12 line}
 
\end{deluxetable}

\begin{deluxetable}{lllllllll}
\tabletypesize{\scriptsize}
\tablecaption{Lines Detected in Orion at 6 GHz.\tablenotemark{1,2,3,4,5} \label{tbl-4}}
\tablewidth{0pt}
\tablehead{
\colhead{Rest Freq(MHz)} & \colhead{Obs Freq(MHz)}  &  \colhead{ID}  &  \colhead{T$_{A}$(K)} & \colhead{T$_{A}$(Corr.)} & \colhead{FWHM(MHz)} & \colhead{kms$^{-1}$(raw)} &  \colhead{kms$^{-1}$(Corr.)} & \colhead{Note}
}

\startdata

5979.13 & 5979.25$\pm0.03$ & H194($\Delta$$n$=7)  &  0.0205 & 0.0205  & 0.368  &  18.4  & 21.6  & -- \\
5981.33 & 5980.94 & NO?  &  0.0087 & 0.0087  &  0.336  &  16.8  & 19.3 &  6 \\
5982.45 & 5982.50$\pm0.041$ & H174($\Delta$$n$=5)  &  0.0547  & 0.0781  &  0.596  &  29.7  & 47.1 & -- \\
5983.30 & --- & H286($\Delta$$n$=24)  & ---  & ---  &  ---  &  ---  & --- & 7 \\
5984.89 & 5984.92 & He174($\Delta$$n$=5) & 0.003  & ---  & ---  & ---  & --- & 8 \\
5985.70 & --- & H257($\Delta$$n$=17) & ---  & ---  &  ---  & ---  &  --- & 9 \\
5987.15 & 5987.18$\pm$.003 & H129($\Delta$$n$=2) & 0.797 & 0.822   &  0.443  &  22.1  &  29.5 & --- \\
--- & 5988.59 & U5988.5 & 0.033   & 0.033  &  0.329   &  16.4  & 18.6 & 10 \\
5989.50 & --- & H266($\Delta$$n$=19) & ---  & ---  &  ---  &  ---  & --- & 11 \\

5989.59 & 5989.67 & He129($\Delta$$n$=2) & 0.0821 & 0.106 & 0.562 & 28.0 & 43.7 & \\
5991.18 & 5990.54? & H224($\Delta$$n$=11)  & $\sim0.004$  & --- & ---  &  --- & --- & 12 \\

5993.35 & 5992.99 & NO? & 0.0084  & 0.0084  &  0.373  &  18.6  & 22.1 & 13 \\
5994.48 & 5993.89$\pm.060$ & H247($\Delta$$n$=15) & 0.0039  & ---  &  ---  & ---  & --- & 14 \\
--- & 5994.59 & U5994.6 & 0.0120   & 0.0120   &  0.375  &  18.7  & 22.2 &  15 \\
5997.79 & 5997.43$\pm.05$ & H252($\Delta$$n$=16) & 0.0066  & 0.0068  &  0.456  &   22.7  &  30.7 &  \\
6000.77 & --- & H282($\Delta$$n$=23) & $<0.002$  & ---  &  ---  &  ---  &  ---  & 16 \\

6002.22 & 6002.26$\pm.031$ & H210($\Delta$$n$=9) & 0.0124 & 0.0127 & 0.425 & 21.2 & 27.5 & \\
6003.45 & 6003.35$\pm0.038$ & H236($\Delta$$n$=13)  & 0.0066  & 0.0068 & 0.355  &  17.7 & 20.8 & \\
6005.63 & $\sim6005$ & $^{17}$OH$^{2}\Pi_{3/2}$  & ---  & --- & ---  &  --- &   --- & 17 \\
6006.66 & 6006.80 & H270($\Delta$$n$=20)  & ---  & --- & ---  &  --- & --- & 18 \\
--- & 6009.03 & U6009.0 & 0.0078  & 0.0098  &  0.693  &  28.0  & 43.1 & 19 \\
6011.14 & --- & H278($\Delta$$n$=22) & ---  & ---  &  ---  & ---  & --- & 20 \\
6011.40 & 6011.36$\pm.050$ & H230($\Delta$$n$=12) & 0.005   & 0.005   &  0.43  &  19.0  & 22.9 &  21 \\
6013.47 & 6013.04$\pm.11$ & H274($\Delta$$n$=21) & 0.0016  & 0.0016  &  0.464  &   21.5  &  28.2 &  \\

6016.75 & 6016.24 & OH$^{2}\Pi_{3/2}$ & 0.0018  & 0.0018  &  0.40  &  20.0  & 25.0  & 22 \\
6016.59 & 6016.91$\pm.11$ & H217($\Delta$$n$=10) & 0.0045 & 0.0048 & --- & --- & --- & 23 \\
6019.23 & 6018.08$\pm.10$ & H289($\Delta$$n$=25)  & 0.0006  & 0.0007 & 0.40  &  20.0 & 25.0 & 24 \\
--- & 6021.56 & U6021.5  & 0.0043  & 0.0046 & 0.485  & 34.1 & 56.8 & \\
6022.61 & 6022.74$\pm.038$ & H202($\Delta$$n$=8) & 0.0127  & 0.0163  &  0.561  &  28.0  &  43.4  &  \\
6025.58 & 6025.62$\pm.003$ & H147($\Delta$$n$=3) & 0.2903 & 0.302 & 0.452 & 22.6 & 30.5 & \\
6027.20 & 6027.06 & H261($\Delta$$n$=18)  & 0.0103  & 0.0104 & 0.394  &  19.6 & 24.2 & 25 \\
6028.04 & 6028.11 & He147($\Delta$$n$=3)  & 0.0302  & 0.050 & 0.622  &  31.0 & 50.8 & \\
6030.75 & 6030.29 & OH$^{2}\Pi_{3/2}$ & 0.0176 & 0.0176  &  0.203  &  10.1  &  10.1  & 26 \\
6035.09 & 6034.65 & OH$^{2}\Pi_{3/2}$ & 0.0434 & 0.0434 & 0.211 & 10.5 & 10.5 & 27 \\
6035.27 & --- & OH?  & 0.0103  & 0.0103 & 0.104  &  5.81 & 5.8 & \\
6036.96 & 6037.04$\pm.02$ & H184($\Delta$$n$=6)  & 0.0339  & 0.0413 & 0.546  &  27.2 & 41.2 & \\
6039.43 & 6039.61 & He184($\Delta$$n$=6) & 0.0045  & 0.0082  & 0.652  &  32.5  &  55.1  &  \\

--- & 6042.4 & U6042.4 & 0.0025 & 0.0025 & 0.142 & 7.1 & 7.2 & \\
6045.52 & 6045.50$\pm.070$ & H241($\Delta$$n$=14)  & 0.003  & 0.003 & 0.39  &  19.5 & 24.2 & \\
6106.86 & 6106.89$\pm.0008$ & H102($\Delta$$n$=1)  & 3.588  & 3.661 & 0.422  &  21.05 & 27.3 & \\
6109.35 & 6109.39 & He102($\Delta$$n$=1)  & 0.408  & 0.429 & 0.469  &  23.4 & 32.0 & \\

\enddata 

\end{deluxetable}

\begin{deluxetable}{ll}
\tabletypesize{\scriptsize}
\tablecaption{Notes to Table 4. \label{tbl-5}}
\tablewidth{0pt}

\tablehead{
\colhead{Note number} & \colhead{Note}
}

\startdata

1 & Offset width = 8 channels = 0.624 MHz = 31.15 kms$^{-1}$ \\
2 & assumes $\Delta$V(Dopp)=27.8 kms$^{-1}$ for H(102,1) \\
3 & FWHM assumed for He(102,1) is 27.8 kms$^{-1}$ \\
4 & 1$\sigma$ rms noise $<0.001$ for 5979 $< \nu < 6005$ MHz  \\
5 & 1$\sigma$ rms noise $<0.0005$ for 6005 $< \nu < 6050$ MHz \\
6 & NO at V$_{LSR}$ of +19 kms$^{-1}$ \\
7 & in wing of H(174,5) line \\
8 & in wing of H(129,2) line \\
9 & in wing of H(129,2) line \\
10 & Deuterium? H$_{2}$? \\
11 & confused with He(129,2) line \\ 
12 & in wing of He(129,2) line (interference?) \\ 
13 & NO at LSR vel of +18 kms$^{-1}$ \\
14 & not cleaned (confused) \\
15 & too strong for H(247,15) line \\ 
16 & not cleaned \\
17 & likely interference \\ 
18 & not cleaned \\
19 & $^{34}$SO$_{2}$? CH$_{4}$? \\ 
20 & confused with H(230,12) \\
21 & no fit made, edge of window \\ 
22 & confused with H(217,10) \\
23 & confused with OH \\
24 & 1$\sigma$ feature \\
25 & Deuterium? $^{3}$He? \\
26 & at V$_{LSR}$ of 23 kms$^{-1}$ \\ 
27& at V$_{LSR}$ of 22 kms$^{-1}$ \\
 
\enddata
\end{deluxetable}

\begin{deluxetable}{llllllll}
\tabletypesize{\scriptsize}
\tablecaption{Hydrogen Recombination Lines in W51\tablenotemark{1} between 5975 and 6050 MHz with $\Delta$$n < 20$\tablenotemark{2}. \label{tbl-6}}
\tablewidth{0pt}

\tablehead{
\colhead{Frequency(MHz)} & \colhead{$\Delta$$n$}  &  \colhead{$n$}  &  \colhead{T$_{A}$(mK)($\pm{0.5}$)(Raw)} & \colhead{(Corr.)} & \colhead{FWHM(MHz)} & \colhead{kms$^{-1}$(raw)} &  \colhead{kms$^{-1}$(Corr.}
}

\startdata

6106.856 & 1 & 102  &  1174.0 & 1235.4  & 0.450  &  22.10  & 31.5 \\
5987.147 & 2 & 129  &  235.72 & 250.8  &  0.452  &  22.64  & 32.2 \\
6025.584 & 3 & 147  &  95.15  & 101.2  &  0.454  &  22.60  & 32.1 \\
5982.446 & 5 & 174  & 25.64  & 29.1  &  0.495  &  24.82  & 37.5 \\
6036.964 & 6 & 184 & 14.24  & 16.2  &  0.495  &  24.59  & 37.0 \\
5979.132 & 7 & 194 & 9.70  & 11.5  &  0.517  &  25.93  &  40.1 \\
6022.606 & 8 & 202 & 7.15   & 8.6   &  0.523  &  26.04  &  40.6 \\
6002.216 & 9 & 210 & 5.77   & 6.6  &  0.495   &   24.73  &  37.4 \\
6016.590 & 10 & 217 & ---  & ---  &  ---  &  ---  &  confused with OH  \\
5991.178 & 11 & 224 & 3.75 & 4.4 & 0.506 & 25.33 & 38.4 \\
6011.398 & 12 & 230  & 2.93  & 3.1 & 0.436  &  21.75 & 30.6 \\
6003.451 & 13 & 236 & 2.14  & 2.1  &  0.305  &  15.24  & 17.7 \\
6045.523 & 14 & 241 & 3.19  & 3.2  &  0.334  &  16.57  &  20.1 \\

5994.475 & 15 & 247 & ---   & ---   &  ---  &  ---  & may contain interference \\
5997.790 & 16 & 252 & 1.23  & ---  &  0.546  &   27.31  &  confused \\
5985.696 & 17 & 257 & ---  & ---  &  ---  &  ---  &  in clean area of $\Delta$$n$=2  \\
6027.198 & 18 & 261 & --- & --- & --- & --- & in clean area of $\Delta$$n$= 3 \\
5989.503 & 19 & 266  & ---  & --- & ---  &  --- & confused with He$\Delta$$n$=2 \\
6006.661 & 20 & 270  & ---  & --- & ---  &  --- & close to interference \\

\enddata 

\tablenotetext{1}{assumes $\Delta$V(Doppler) = 31.5 km s$^{-1}$}
\tablenotetext{2}{offset width = 31.15 km s$^{-1}$}
 
\end{deluxetable}

\begin{deluxetable}{lllllll}
\tabletypesize{\scriptsize}
\tablecaption{Helium Recombination Lines in W51 Between 5975 and 6050 MHz with $\Delta$$n < 9$. \label{tbl-7}}
\tablewidth{0pt}

\tablehead{
\colhead{Frequency(MHz)} & \colhead{$\Delta$$n$}  &  \colhead{$n$}  &  \colhead{T$_{A}$(mK)} & \colhead{Rel. Line Area} & \colhead{FWHM(MHz)} & \colhead{comments}
}

\startdata

5981.573 & 7 & 194  &  --- & ---  & ---  &  in clean area of H(174,5) \\
5984.888 & 5 & 174  &  1.69 & 0.28  & 0.156 &  in clean area of H(129,2) \\
5989.591 & 2 & 129  & 17.06  & 8.45  &   0.465  &  ---  \\
6025.065 & 8 & 202 & --- & ---  &  ---  &  confused with H(147,3) line \\
6028.044 & 3 & 147  & 6.43  & 2.64  &  0.387  &  ---  \\
6039.429 & 6 & 184  & $<1$  & ---  &  ---  &  ---  \\

\enddata

\end{deluxetable}

\begin{deluxetable}{lll}
\tabletypesize{\scriptsize}
\tablecaption{Other Lines Detected in W51 Near 6.0 GHz. \label{tbl-8}}
\tablewidth{0pt}

\tablehead{
\colhead{Frequency(MHz)}  &  \colhead{T$_{A}$(K)} & \colhead{Identification}
}

\startdata

5992.3 & -0.005 & possible interference  \\
5994.5 & -0.004 & possible interference  \\
5996.7 & 0.002 & U5996.7  \\
5999.0 & 0.003 & U5999.0  \\
6000.0 & 0.002 &  U6000.0   \\
6016.7 & -0.007 & OH   \\
6023.8 & 0.0036 & U6023.8   \\

6030.5 & -0.07 & OH   \\
6030.7 & 0.034 & OH   \\
6035.1 & 0.34 & OH  \\
6042.4 & 0.003 & U6042.4  \\

\enddata

\end{deluxetable}
\clearpage

\begin{figure*}
\hspace{0.0cm}
\vspace{0.0cm}
\includegraphics[width=16cm]{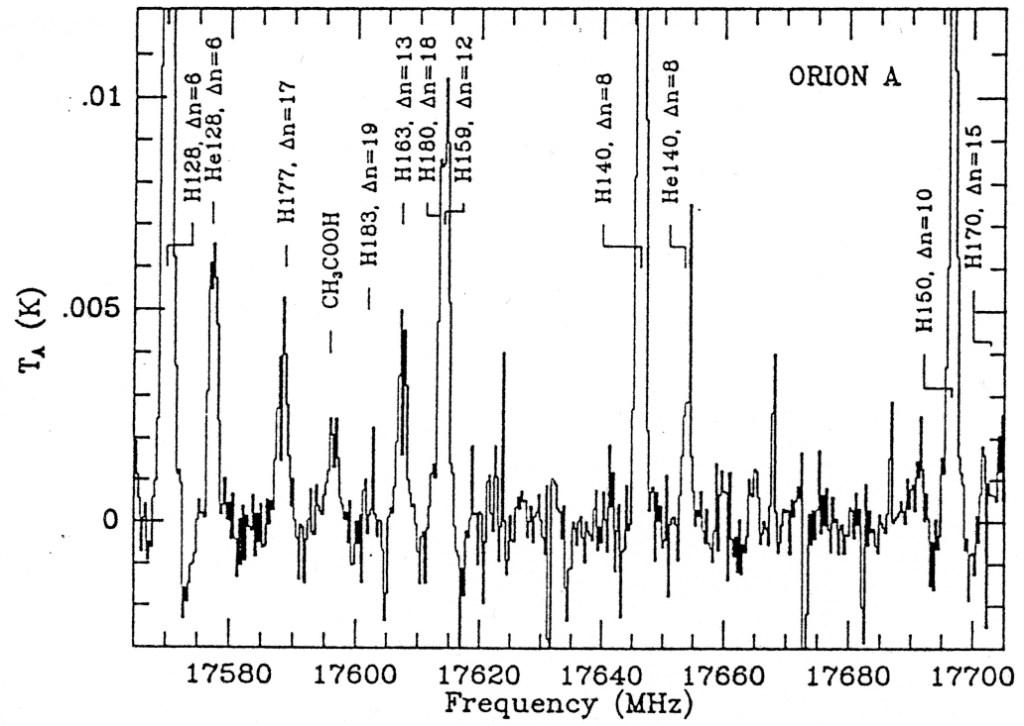}
\caption{{Spectrum of Orion A at 17.6 GHz obtained using frequency switching and presented after reference lines have been "cleaned" off. No sinusoids or polynomials have been removed from the baseline. Locations of the hydrogen and relevant helium lines are indicated. The H(163,13) line appears to be too weak compared to the $\Delta$$n = 12$ and 17 lines and may be confused with a line, or lines, of molecular origin such as the H$_{2}^{+}$ doublet at 17604.3 and 17610.3 MHz (see text). Although an LSR velocity of 8 kms$^{-1}$ was assumed when the data were obtained, the detected frequencies in Table 2 have been adjusted to an LSR velocity of 0 kms$^{-1}$ to agree with the value assumed for the 6 GHz Orion data. \label{fig1}}}

\end{figure*}
 
\begin{figure*}
\hspace{0.0cm}
\vspace{0.0cm}
\includegraphics[width=16cm]{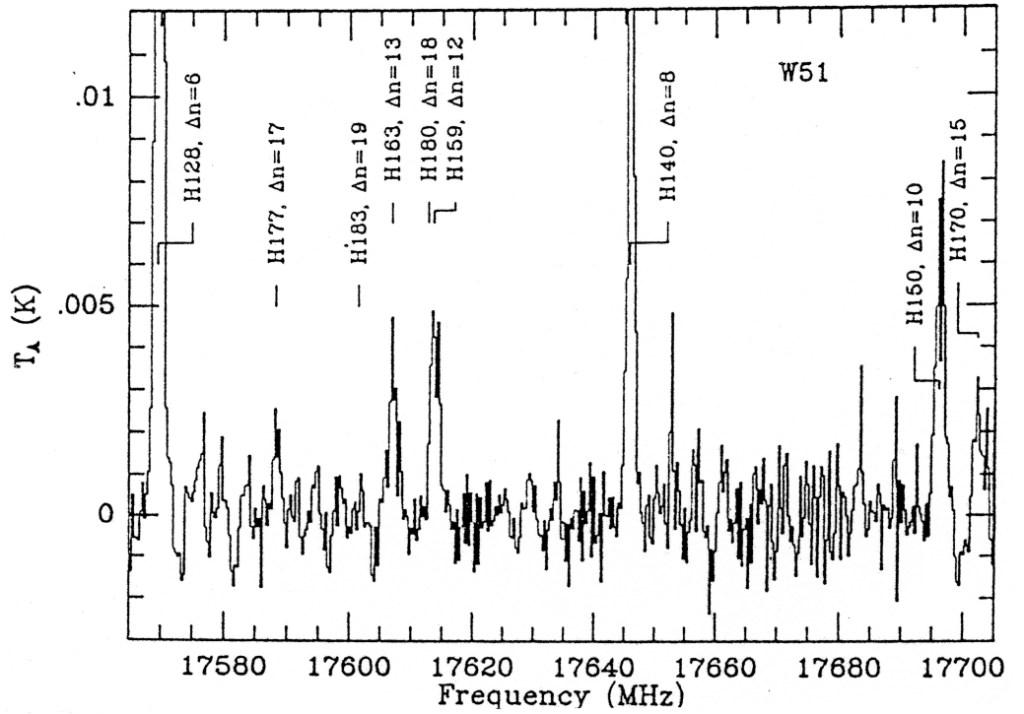}
\caption{{Spectrum of W51 at 17.6 GHz, after frequency-switched reference lines have been removed. The velocity assumed was V$_{LSR}$ = 60 kms$^{-1}$. The slight blueshift in the line position indicates that the true velocity is approximately V$_{LSR}$ = 58 kms$^{-1}$. No sinusoids or polynomials have been removed from the baseline. \label{fig2}}}
\end{figure*}  
 
\begin{figure*}
\hspace{0.0cm}
\vspace{0.0cm}
\includegraphics[width=16cm]{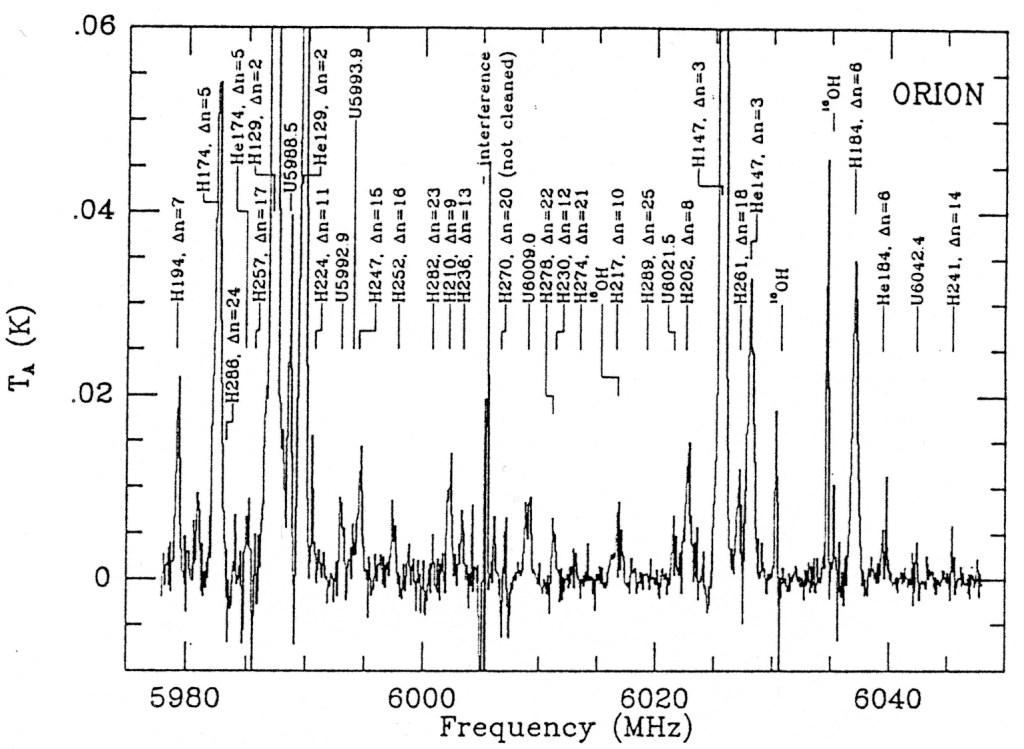}
\caption{{Spectrum of Orion at 6 GHz. No sinusoids or polynomials have been removed. The expected location of recombination lines with $20 < \Delta$$n < 30$ have been indicated for information purposes only and this is not meant to imply that they have been detected. An LSR velocity of 0 kms$^{-1}$ was assumed.\label{fig3}}}
\end{figure*}  
 
\begin{figure*}
\hspace{0.0cm}
\vspace{0.0cm}
\includegraphics[width=16cm]{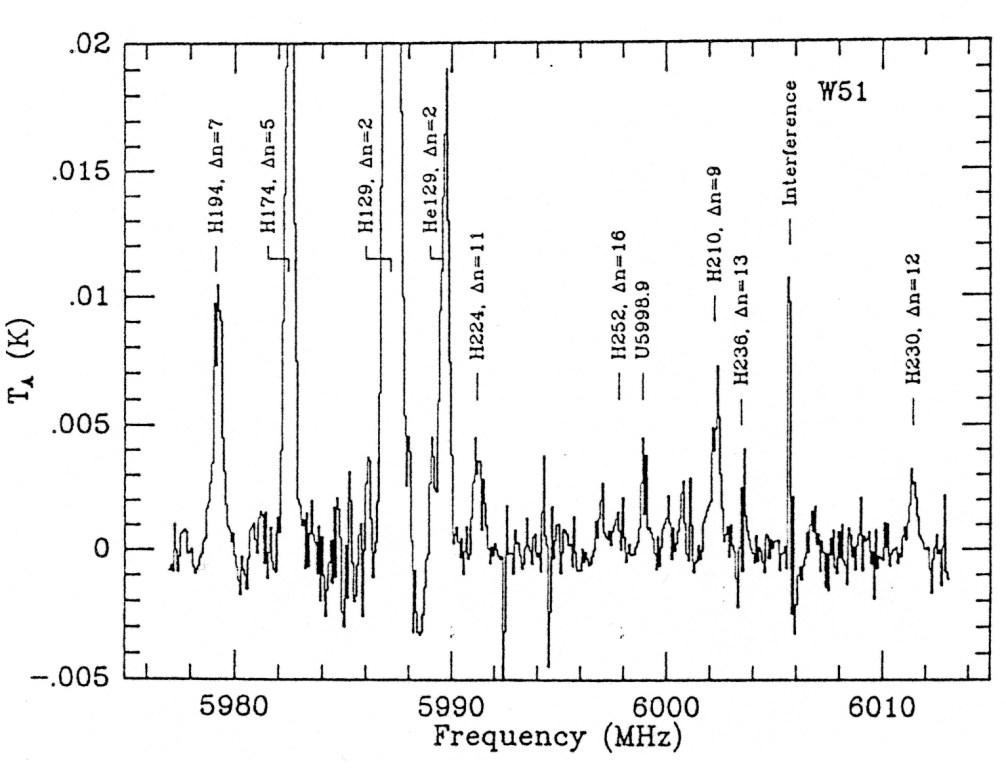}
\caption{{Spectrum of W51 below 6010 GHz showing location of hydrogen and helium recombination lines as well as several molecular and U-lines. The frequency-switched reference lines have been removed but no sinusoids or polynomials have been removed from the baseline. \label{fig4}}}

\end{figure*}

\begin{figure*}
\hspace{0.0cm}
\vspace{0.0cm}
\includegraphics[width=16cm]{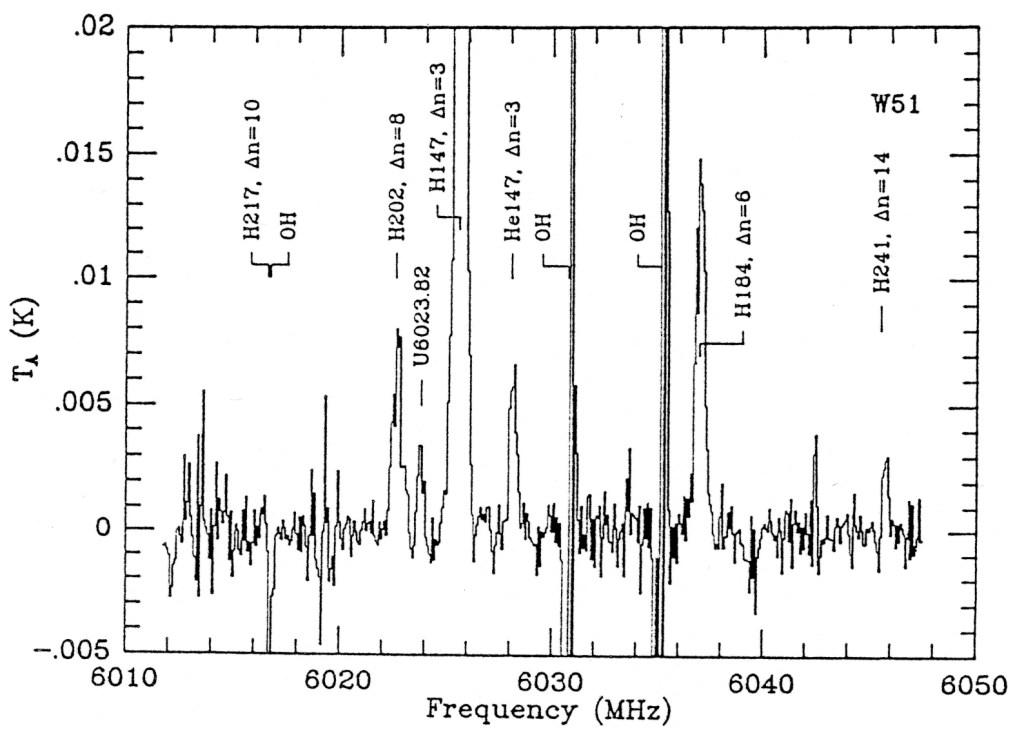}
\caption{{Spectrum of W51 above 6010 MHz. The frequency-switched reference lines have been removed but no sinusoids or polynomials have been removed from the baseline. \label{fig5}}}

\end{figure*}  
 
\begin{figure*}
\hspace{0.0cm}
\vspace{0.0cm}
\includegraphics[width=16cm]{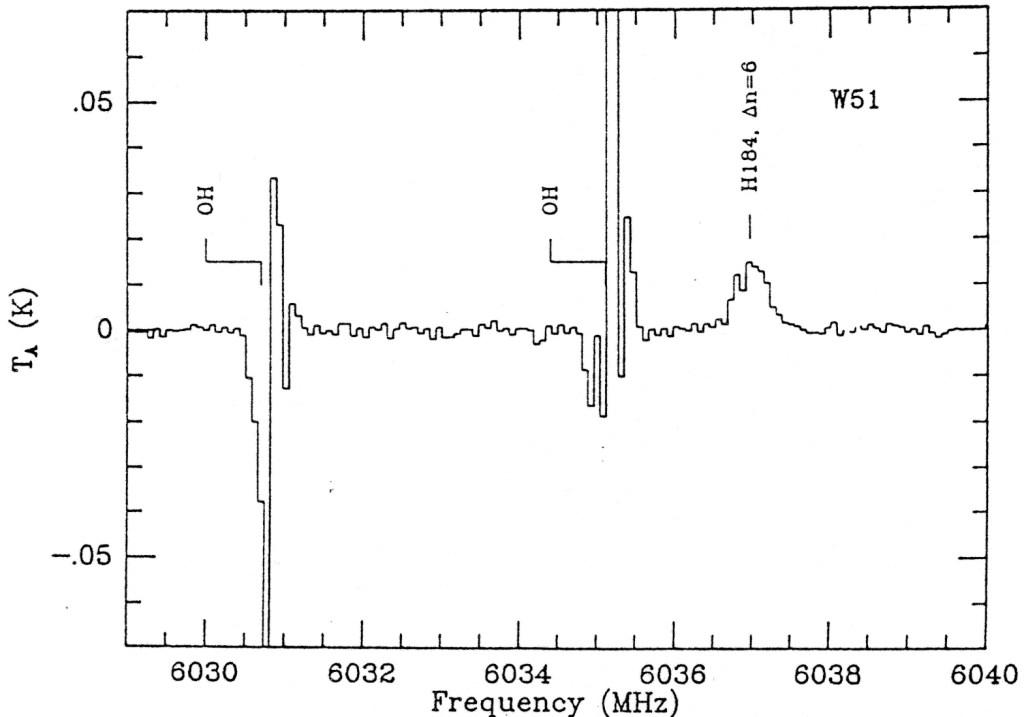}
\caption{{A portion of the spectrum of W51 at 6 GHz showing the two strongest OH lines and the H(184,6) line. Structure is visible in the red wing of all three lines. \label{fig6}}}

\end{figure*}

\begin{figure*}
\hspace{0.0cm}
\vspace{0.0cm}
\includegraphics[width=16cm]{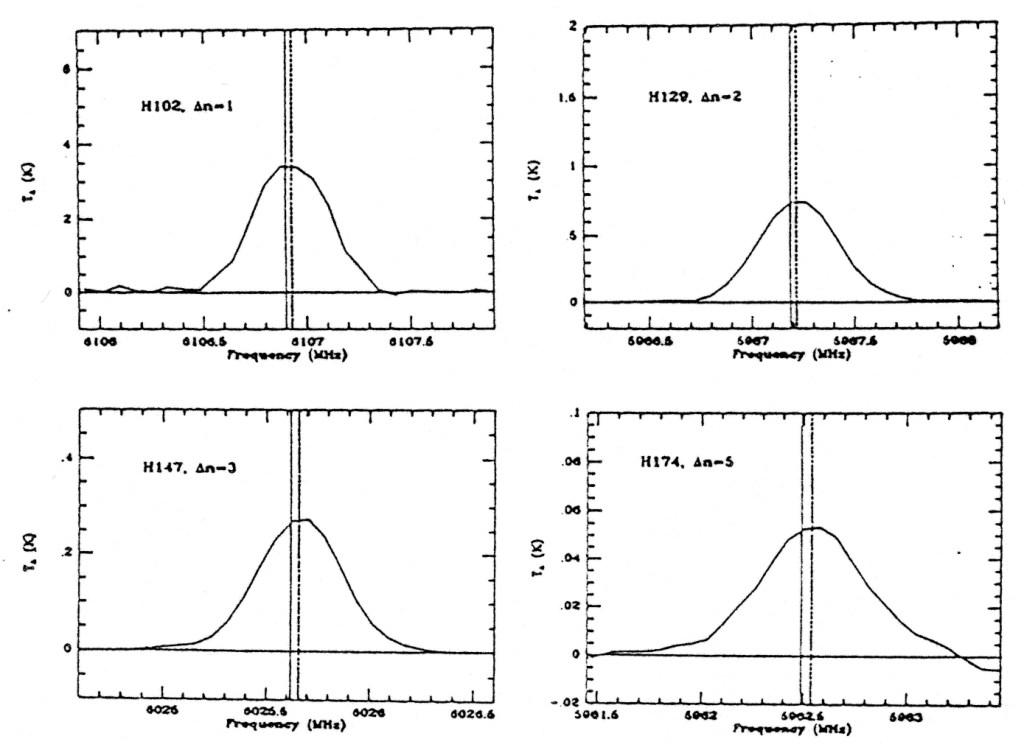}
\caption{{Plots of individual recombination lines observed in Orion at 6 GHz showing the relative position of the line center frequencies. The solid line indicates the predicted position of the line. The dashed line indicates the observed line center. \label{fig7}}}

\end{figure*}  

 \begin{figure*}
\hspace{0.0cm}
\vspace{0.0cm}
\includegraphics[width=16cm]{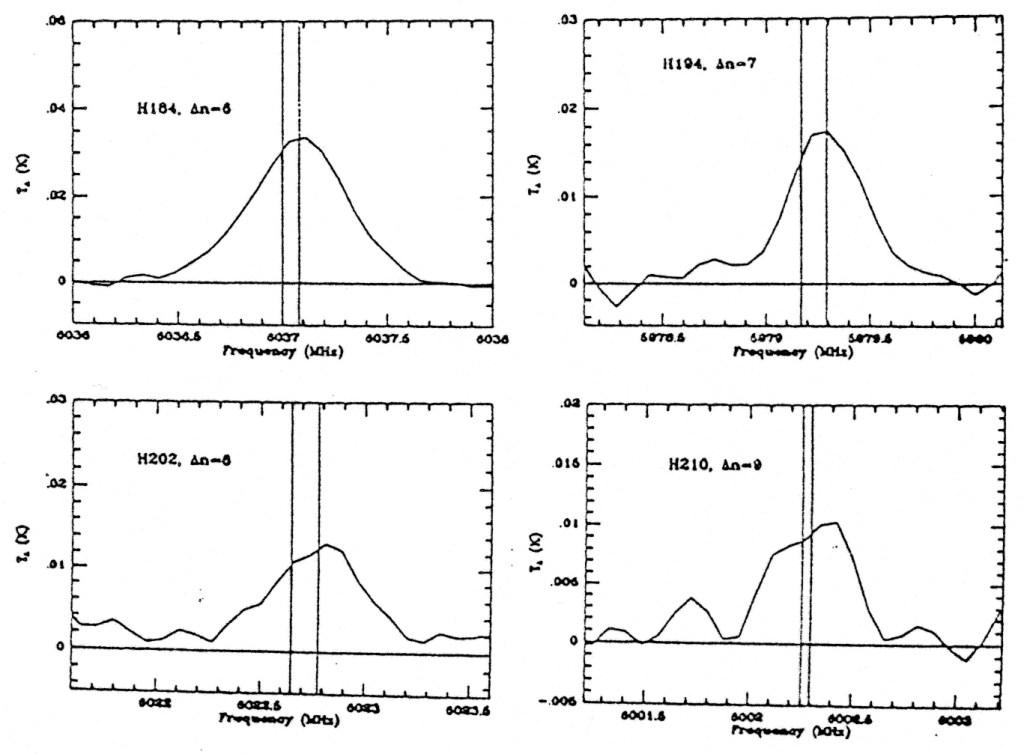}
\caption{{Same as Fig 7. \label{fig8}}}

\end{figure*}  

\begin{figure*}
\hspace{0.0cm}
\vspace{0.0cm}
\includegraphics[width=16cm]{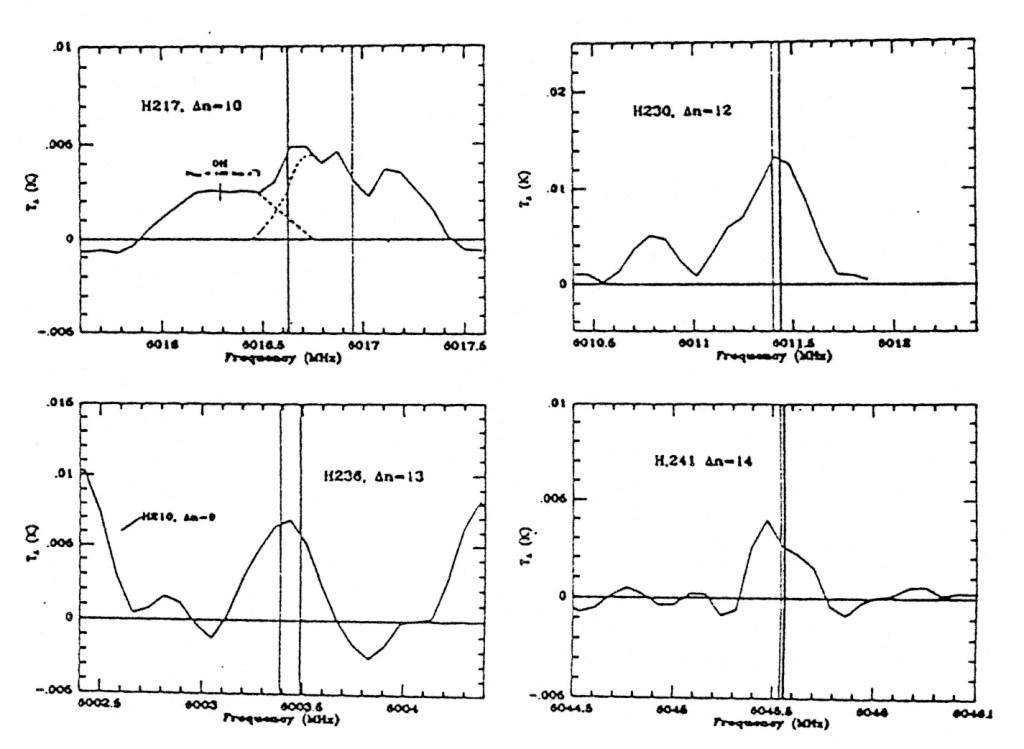}
\caption{{Same as Fig 7. Because the $\Delta$$n$=10 line is confused with an OH line its amplitude and width are uncertain and dashed curves have been drawn to indicate the assumptions made in determining the line position.
\label{fig9}}}

\end{figure*}  

\begin{figure*}
\hspace{0.0cm}
\vspace{0.0cm}
\includegraphics[width=16cm]{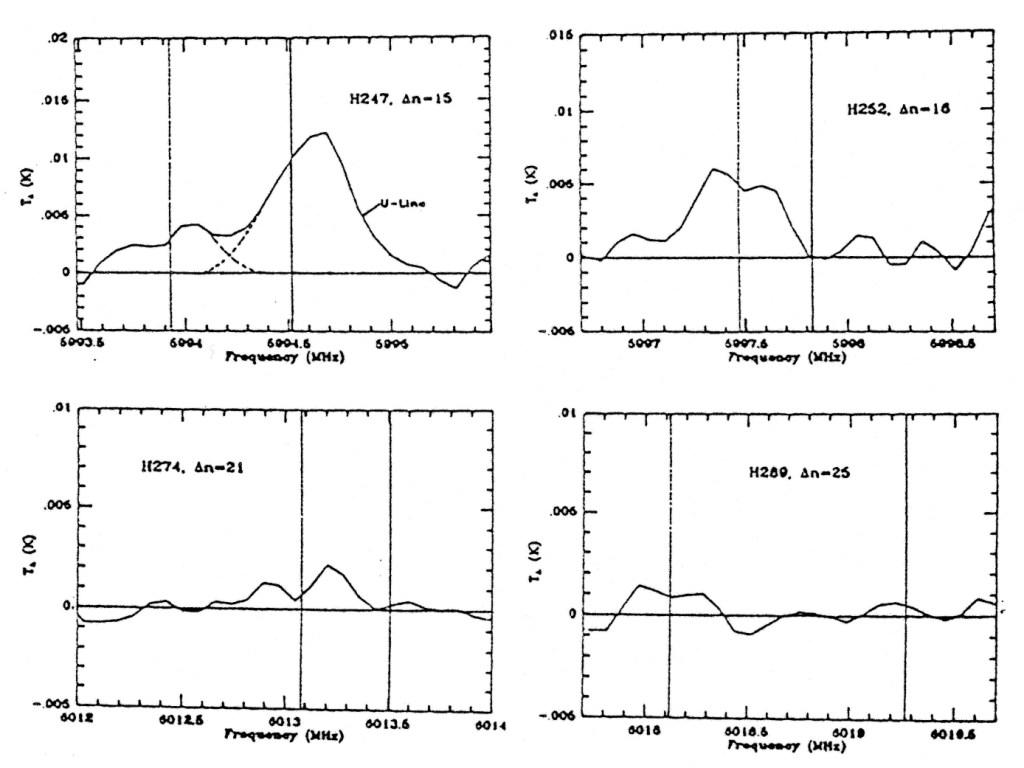}
\caption{{Same as Fig 7. Although the $\Delta$$n$=25 line has been included its identification is questionable. \label{fig10}}}

\end{figure*}  

\clearpage 
\begin{figure}
\hspace{0.0cm}
\vspace{0.0cm}


\includegraphics[width=8cm]{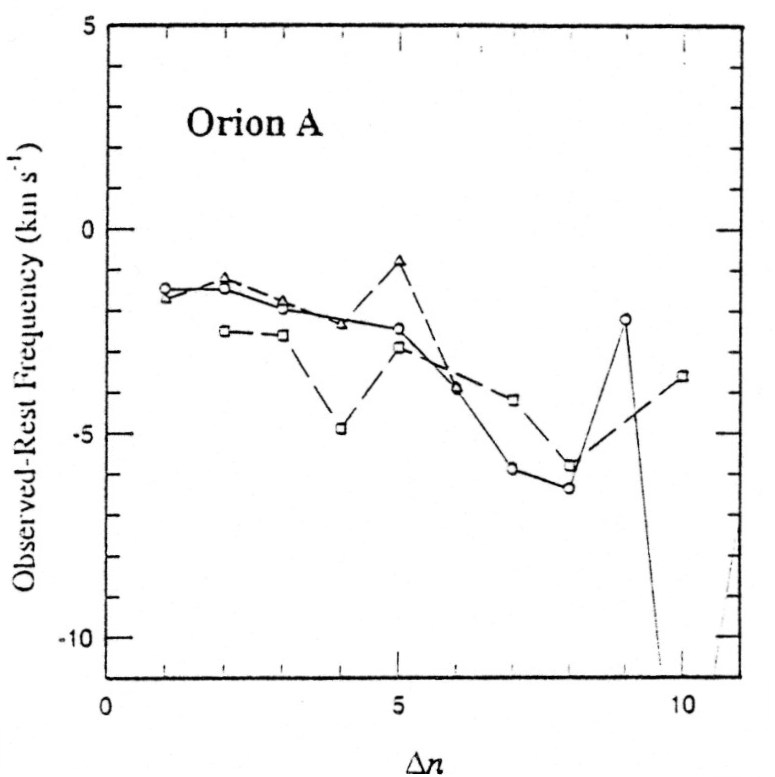}
\caption{{The differences between observed and rest frequencies (expressed in velocity units) for lines near 6 GHz in Orion with $\Delta$$n$-values $< 11$. (circles) data from this paper, (square) data from Rood et al (1984), (triangle) data from Smirnov et al (1984). \label{fig11}}}

\end{figure}




\begin{figure}
\hspace{0.0cm}
\vspace{0.0cm}
\includegraphics[width=8cm]{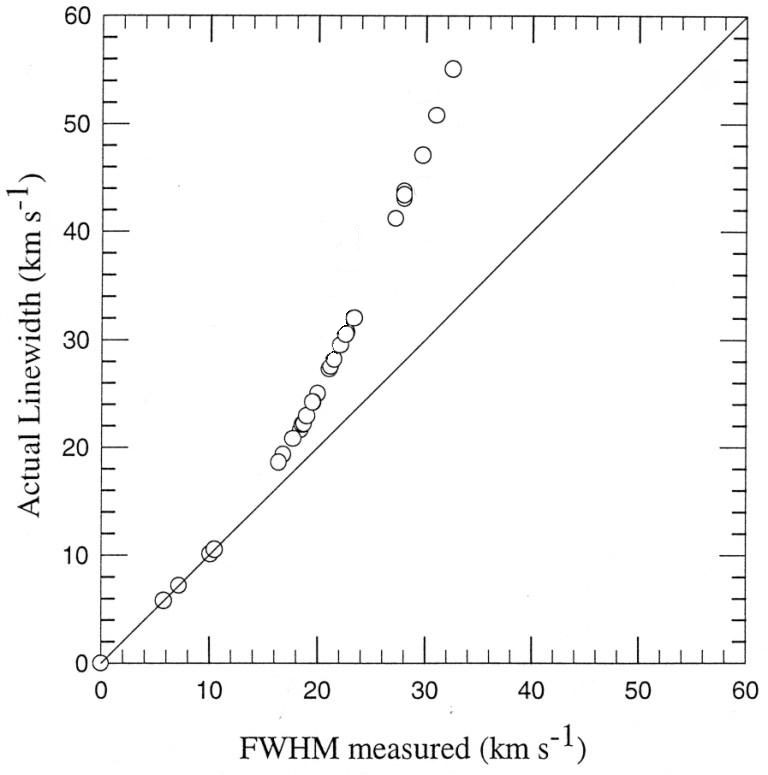}
\caption{{The curve defined by the open circles is used to obtain the true width of a line from the measured linewidth when frequency-switched offsets of $\pm15.5$ kms$^{-1}$ are used. (see Appendix A). \label{fig12}}}

\end{figure}

\begin{figure}
\hspace{0.0cm}
\vspace{0.0cm}
\includegraphics[width=8cm]{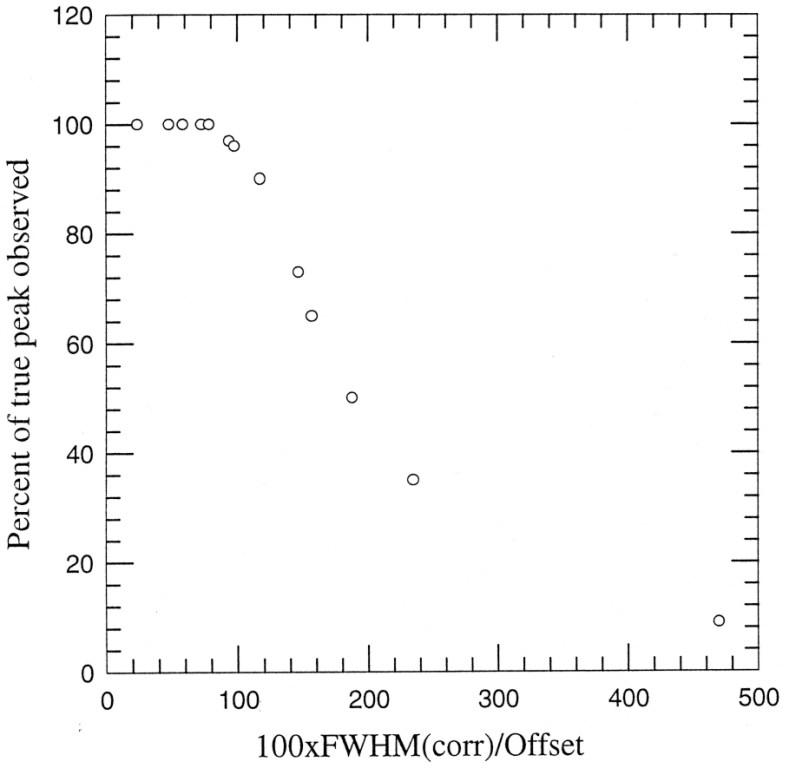}
\caption{{Curve used to obtain the true peak line strength of a wide, frequency-switched line when the true width of the line is known. \label{fig13}}}

\end{figure}

\appendix

APPENDIX A: CORRECTION CURVES TO RECOVER THE TRUE LINE PARAMETERS OF WIDE, FREQUENCY-SWITCHED LINES.

     It is important to have flat baselines for the LINECLEAN program to work effectively and, to achieve this, it was necessary to place the frequency-switched reference line close to the signal line. The width of impact-broadened lines can then be a significant fraction of the frequency-switched interval. When this happens, the measured linewidths, and possibly even the peak values, can be reduced because one wing of the reference line falls on top of the signal line. Correction curves are therefore required so the true line strengths and widths can be recovered.

     To obtain the information needed to correct for this effect, H$\alpha$ lines in Orion A and W51 were observed with increasingly narrower frequency-switched offsets to simulate the effect produced by linewidths that become increasingly more broadened by impact broadening and fill more of the frequency-switched window.  The correction curve appropriate for a frequency-switched offset of $\sim31$ km s$^{-1}$ using the MOR data reduction technique described by \citet{bel97} with 5 overlaps, is given by the open circles in Fig. 12. The method used to obtain this detailed curve is complicated, and will not be discussed here, since it turns out that a perfectly adequate curve can be obtained simply by comparing the measured frequency-switched widths to re-observed, or previously-published widths, obtained on a few sources with different intrinsic linewidths (using position switching or very wide frequency-switched offsets). Although the uncertainty in the curve was expected to increase slightly as the linewidth increased, it was shown by \citet{bel00}
that the corrected linewidths obtained using the curve in Fig 12 agree well with results obtained by others using position switching, even for the widest lines. Furthermore, there appears to be no reason why the curve in Fig. 12 cannot be easily and accurately scaled to fit any frequency-switched offset by comparing the measured linewidths obtained using the new offset to a few known linewidths.

     Once the corrected line widths have been determined, the corrected peak values can be obtained directly from the curve in Fig. 13. This curve is easily obtained by observing any strong line of known width with several different frequency-switched offsets.

     Although the correction curves presented here have been included for illustration purposes, observers wishing to use this technique may wish to derive their own.

\end{document}